\documentclass[12pt]{article}
\usepackage{hyperref}
\usepackage{amssymb}
\usepackage{graphicx}
\usepackage{float}
\usepackage{slashed}
\usepackage{amsmath}
\usepackage{tabularx}
\usepackage{breqn}
\usepackage{authblk}
\usepackage{array}
\usepackage{dsfont}
\usepackage{cite}
\usepackage{physics}
\usepackage[section]{placeins}
\usepackage[a4paper, total={6.8in, 10in}]{geometry}
\newcommand{\be}{\begin{equation}}
\newcommand{\ee}{\end{equation}}
\newcommand{\bea}{\begin{eqnarray}}
\newcommand{\eea}{\end{eqnarray}}
\newcommand{\ba}{\begin{aligned}}
\newcommand{\ea}{\end{aligned}}
\begin{document}
\title{Wormhole Solutions in deformed space-time}

\author{Harsha Sreekumar\thanks{harshasreekumark@gmail.com}, E. Harikumar\thanks{eharikumar@uohyd.ac.in}}
\affil{School of Physics, University of Hyderabad, \\Central University P.O, Hyderabad-500046, Telangana, India}

\maketitle

\begin{abstract}
We construct and analyse wormhole solutions in quantised space-time. The field equations are constructed from the deformed wormhole metric in the proper reference frame using tetrads. The spatial geometry of the wormhole is analysed in kappa space-time. Further, modifications to the conditions that ensure traversibility of the wormhole are studied and it is found that the necessity of the exotic matter persists in the non-commuative case as in the commutative space-time. Casimir energy is considered a possible source for exotic matter and it is shown that the time to pass through the wormhole as well as the amount of exotic matter required to create the wormhole reduce due to non-commutativity of space-time. \\

\end{abstract}

\section{Introduction}

Einstein's General Relativity is a well-established theory of gravitation where force of gravity is attributed to the geometrical effect on space-time imparted by matter sources \cite{gravitationthorne}. It is an experimentally well-established theory and most of its predictions like gravitational waves, black holes, etc have been verified \cite{abbott, akiyama}. One such theoretical prediction from relativity is wormhole which allows travel between two different, far off asymptotically flat regions of space-time of the universe by connecting them through a throat. It is a solution of Einstein's field equation and it was shown that the Schwarzschild solution also represent a wormhole \cite{Flamm}. In 1935, Einstein and Rosen proposed the Einstein-Rosen bridge that connects two identical sheets of space-time through a \lq\lq bridge" \cite{einsteinrosen}. The notion of traversible wormhole was first given by Morris and Thorne in 1988 where further conditions were imposed on the metric variables and field equations to ensure traversibility. The wormhole metric is given by \cite{kipthorne}, 
\be \label{wormhole}
ds^2=-e^{2\Phi(r)}c^2dt^2+\frac{dr^{2}}{\left(1-\frac{b(r)}{r}\right)}+r^2d\theta^{2}+r^2sin^{2}\theta d\phi^{2},
\ee
where the redshift function, $\Phi(r)$ and the shape function, $b(r)$ depends only on $r$. Traversibility of the wormhole requires the solution to have a throat which has a minimum radius and connects two different regions of the universe, absence of horizon, acceleration and tidal acceleration felt by the traveller to be of the order of $1g_{earth}$ and the travel time of the traveller to be sufficiently small. Existence of wormhole curvature necessitates the requirement of exotic matter which is defined by the property that its tension exceeds energy density. It was shown that exotic matter violates weak energy condition and so does Casimir energy, hence it can be a candidate for such exotic matter \cite{morristhorne}. Such wormhole solutions are called Casimir wormholes. Casimir energy is the energy produced by vacuum fluctuations of electromagnetic fields when two conducting surfaces are present close to each other.

Although experimental realization of wormhole is unaccomplished, various studies are ongoing in this field. Some examples of traversible wormhole by discarding spherical symmetry and confining the exotic matter to a very small region such that the traveller will not have to encounter this region in his or her travel was presented in \cite{visser}.  Rotating wormholes were developed in \cite{teo} where the traveller can avoid regions of exotic matter confined to the throat. Such wormholes were also found to have ergoregions from which one can remove energy via Penrose process. Effect of wormholes on quantum fields in a Euclidean space at low energy limits is studied in \cite{hawking}. Here effective interactions are incorporated in flat space-time that create or annihilate closed universes that contain a particular number of particles. Consequences of Casimir energy on a traversible wormhole was studied in \cite{garattini}. In \cite{sorge}, a Casimir apparatus is considered to orbit an Ellis wormhole in the equatorial plane and it was found that for an observer orbiting with the apparatus, the Casimir energy density reduces. Such an analysis of a rotating Casimir apparatus around a Schwarzschild and rotating wormhole have also been done in \cite{santoz, muniz}. In \cite{shingo}, phase structure of  transitions from
normal to BEC states in the traversable wormhole is analysed and possibility of existence of Josephson junction near the wormhole throat is argued. Violation of null energy condition in a spherically symmetric and static  traversable was studied in \cite{battista}. In \cite{battista2, falco, falco2, falco3}, wormhole solutions in different modified gravity models were analysed. Geometry of throat in the \textit{k}-essence setting is reported in \cite{carlos}.

Understanding the nature of gravity at microscopic scales compels one to go beyond general theory of relativity. Various studies in search of such a theory are being pursued and one such paradigm is non-commutative(NC) space-time. A fundamental length scale taken to be of the order of Planck length appears in all models of quantum gravity which appears naturally in NC space-time\cite{glikman, connes, dop}. NC space-time is also found to emerge in the limit of low energy in some models of quantum gravity like string theory etc \cite{doug, livine}. These features makes it an interesting model to study various physical phenomena. Casimir wormhole geometry have been studied in NC geometry in \cite{kuhfittig}. The non-commutativity is introduced here through the smearing of the Casimir energy density which forms component of the stress tensor while the Einstein tensor is left unaffected by non-commutativity and the resulting wormhole has a throat that is a smeared surface. In \cite{sahoo2}, wormhole solutions are studied in $f(R, L_{m})$ gravity in the background of NC geometry. Here the Lorentzian and Gaussian form of NC correction coming from the Moyal space-time parameter $\theta$ to the Casimir energy is considered. Also the shape function of the wormhole that complies with the throat condition is derived and the behaviour of the shape function with model dependent parameters are studied. The stability of the wormhole was also found to be dependent on the model parameters. \cite{pradhan} studies wormhole geometry in Rastall- Rainbow gravity in which non-commutativity is brought in by Gaussian and Lorentzian distribution of energy density in the wormhole metric. Rainbow gravity emerges by the generalisation of doubly special relativity to curved space-time whose geometry changes with the energy of the observer probing the space-time \cite{rainbow}. Rastall theory modifies the conservation laws where the covarient derivative of the energy-momentum is dependent on the derivative of the Ricci scalar \cite{rastall}. Violation of null energy condition in radial and lateral directions are found to depend on the model parameters and it was seen that in case of Gaussian distribution the null energy condition is violated irrespective of the model parameter \cite{pradhan}.

Various studies of wormhole solutions have been reported in modified theories of gravity. Wormhole solutions are analysed in f(R) gravity in \cite{mishra}. Here, three models by choosing three different forms of the shape function are studied and the violation of null energy condition for all the three are analysed. It is shown that the null energy condition is violated only below a critical radius in one of the models and in another it failed to violate the null energy condition for tangential pressure. In \cite{azizi}, it is shown that in f(R, T) gravity, presence of exotic matter is not necessary to support the wormhole geometry, rather this can be done by an effective stress-energy tensor obtained as a result of extra terms of curvature and matter. Necessity of exotic matter is eliminated in \cite{lobo} where traversible wormhole solutions in f(R) gravity theories were analysed. Here the wormhole stress-energy tensor satisfy the energy conditions but the wormhole geometry is supported by the higher order terms coming from curavture (called as gravitational fluid). In \cite{sahoo}, Casimir wormholes are analysed in teleparallel gravity and the behaviour of the shape function is analysed for different configures of the Casimir apparatus.

In this paper, we analyse the impact of space-time non-commutativity on traversible wormhole solutions. After constructing the $\kappa$-deformed wormhole metric, the deformed Einstein equation is derived using tetrads and the effect of space-time $\kappa$-deformation on the geometry of the wormhole is studied. Further, we look into the impact of non-commutativity on the conditions necessary for the wormhole to be traversible, such as total time for the travel through the wormhole, acceleration felt by the traveller and tidal acceleration felt by the traveller. Necessity of exotic matter for the existence of wormhole is demonstrated in NC space-time as well. With Casimir energy used as exotic matter, the constraints imposed for travel through a deformed wormhole is studied. In Section $2$, we give a brief introduction on kappa space-time and the construction of the deformed metric in the same. In section $3$, field equations for the wormhole is constructed by transforming the necessary tensors to a proper reference frame using tetrads. It also requires generalization of the energy-momentum tensor to the deformed space-time. Section $4$ analyses the effect of non-commutativity on the conditions necessary for a wormhole to be traversible and section $5$ examines the effects of these constraints on the components of the stress-energy tensor via the field equations. Section $6$ looks into the effect of Casimir energy on all the constraints of a traversibile wormhole in NC space-time. We present our conclusions in Section $7$. Appendix A discusses the construction of $\kappa$-deformed Minkowski metric. Construction of $\kappa$-deformed Lorentz transformtaion is given in Appendix B. The boundary conditions of a $\kappa$-deformed wormhole is discussed in Appendix C and Appendix D discusses the $\kappa$-deformed wormhole solution when the exotic matter is confined to a very small region around the throat.

\section{$\kappa$-deformed space-time}

Experiments concord with the assumption that space-time is a differential manifold and all fundamental theories are developed as field theories on a differential manifold. But various developments to understand the nature of gravity at extremely small length scales led to discrete models of space-times. Naturally this opened up analysis of large number of physical phenomena in such space-times. $\kappa$-deformed space-time is such a NC space-time whose space coordinates commute whereas the space and time coordinates do not, i.e.,
\be
[\hat{x}^i,\hat{x}^j]=0,~~~[\hat{x}^0, \hat{x}^i]=ia\hat{x}^i,~~~a=\frac{1}{\kappa}. \label{ksp-1}
\ee
Here $a$ is the deformation parameter with \textit{length} dimension. This space-time is associated with different quantum gravity model as well as with deformed special relativity \cite{glikman, livine}. Field theory models were developed in $\kappa$-deformed space-time by replacing the pointwise multiplication between the fields by star product and they are invariant under $\kappa$-Poincare transformation \cite{lukieriski}. Field theoretical models can also be studied in $\kappa$-deformed space-time by re-writing $\hat{x}^{\mu}$ in terms of $x^{\mu}$ and their derivatives. Various realizations of $\kappa$-Minkowski space-time coordinates are known to be on an equal footing in terms of the physical meaning rendered to the phenomena studied \cite{meljanac}.

Here, realization approach where the NC coordinate, $\hat{x}^{\mu}$ are expressed as \cite{meljanac}, 
\be \label{ksp-2}
 \hat{x}_0=x_0\psi(ap^{0})+iax_j\partial_j\gamma(ap^{0});~~~~\hat{x}_i=x_i\varphi(ap^{0}),
\ee
is used. Compatibility between eq.(\ref{ksp-2}) and eq.(\ref{ksp-1}) impose restrictions on $\psi(ap^{0}), \gamma(ap^{0})$ and $\varphi(ap^{0})$ apart from the requirement that eq.(\ref{ksp-2}) and eq.(\ref{ksp-1}), in the limit $a \rightarrow 0$ should reduce to the usual Minkowski result \cite{meljanac}. Choosing $\psi(ap^{0})=1$ allowed by above constraints, we find that 
\be \label{ksp-3}
 \hat{x}_0=x_0+iax_j\partial_j\gamma(ap^{0});~~~~\hat{x}_i=x_i\varphi(ap^{0}).
\ee
Further, we choose $\varphi=e^{-ap^{0}}$ for our study and the free particle dispersion relation in $\kappa$ space-time takes the form,
\be
\frac{4}{a^2}\sinh^2 \bigg(\frac{ap^{0}}{2}\bigg) -p_ip_i ~e^{ap^{0}}-m^2c^2 +\frac{a^2}{4}\left[\frac{4}{a^2}\sinh^2\bigg(\frac{ap^{0}}{2}\bigg) -p_ip_i~ e^{ap^{0}}\right]^2=0,\label{disp}
\ee
where $p_{i}$ is the particle three momentum and $p^{0}$ is the energy of the particle probing the space-time. Keeping terms only upto first order in the deformation parameter $a$, eq.(\ref{disp}) becomes
\be 
\hat{E}=E g(E,a),
\label{disp2}
\ee
where 
\be 
g(E,a)=\left[1+\frac{ap^{0}}{2}\left(1-\left(\frac{mc^2}{E}\right)^2\right)\right].
\ee
The generalization of space-time metric to kappa-deformed space-time is given as \cite{sir, zuhair},
\be \label{deformedlineelement}
 \hat{g}_{\mu\nu}=g_{\alpha\beta}(\hat{y})\Big(p^{\beta}\frac{\partial \varphi^{\alpha}_{\nu}}{\partial p^{\sigma}}\varphi_{\mu}^{\sigma}+\varphi_{\mu}^{\beta}\varphi_{\nu}^{\alpha}\Big),
 \ee
where $\hat{y}_{0}=x_{0}-ax_{j}p^{j}$ and $\hat{y}_{i}=x_{i}$. Also $\varphi_{\mu}^{\nu}=\delta_{\mu}^{\nu}e^{-ap^{0}}$. This gives the generic form of line element in kappa space-time to be,
\be \label{deformedmetric}
\begin{aligned}
d\hat{s}^2&=g_{00}(\hat{y})dx^0dx^0+\Big(g_{i0}(\hat{y})\big(1-ap^0\big)-ag_{im}(\hat{y})p^m\Big)e^{-2ap^0}dx^0dx^i\\&+g_{0i}(\hat{y})e^{-2ap^0}dx^idx^0+g_{ji}(\hat{y})e^{-4ap^0}dx^idx^j.
\end{aligned}
\ee
Using eq.(\ref{deformedmetric}) and eq.(\ref{wormhole}), we find the $\kappa$-deformed wormhole metric as,
\be \label{deformedwormhole}
d\hat{s}^2=-e^{2\Phi(r)}c^2dt^2+e^{-4ap^{0}}\left[\frac{dr^{2}}{\left(1-\frac{b(r)}{r}\right)}+r^2d\theta^{2}+r^2sin^{2}\theta d\phi^{2}\right].
\ee

\section{Deformed Einstein's field equations in the wormhole background}

In this section, we construct the field equations and by inverting these we obtain constraints that the components of the energy-momentum tensor need to satisfy for the wormhole to be traversible. The corresponding Christoffel symbols and Reimann tensors are constructed for the $\kappa$-deformed wormhole metric (eq.(\ref{deformedwormhole})) in the $(ct, r, \theta, \phi)$ coordinate. The basis vectors of this coordinate system are $(\hat{e}_{t},\hat{e}_{r},\hat{e}_{\theta},\hat{e}_{\phi})$. For the ease of physical interpretation and calculations, we shift to a \lq\lq proper reference frame" with an orthonormal set of basis vectors where the observers are at rest \cite{kipthorne}. This is done using tetrads that are generalized to NC space-time using $\kappa$-deformed Minkowski metric (see Appendix-A for details). We have,
\be \label{metricgeneral}
d\hat{s}^2=\hat{g}_{\mu\nu}~d\hat{x}^{\mu}d\hat{x}^{\nu}=\hat{\eta}_{ab}~\hat{e}^{\tilde{a}} \otimes \hat{e}^{\tilde{b}},
\ee
where $\hat{g}_{\mu\nu}$ is given in eq.(\ref{deformedwormhole}) and $\hat{\eta}_{ab}$ in eq.(\ref{deformedminkowski}).
This gives the deformed basis vectors in the proper reference frame of coordinates $(c\tilde{t},\tilde{r},\tilde{\theta},\tilde{\phi})$ as,
\be \label{properbasis}
\hat{e}^{\tilde{t}}=e^{\Phi}cdt;~~~\hat{e}^{\tilde{r}}=e^{-ap^{0}}\frac{dr}{\sqrt{1-\frac{b(r)}{r}}};~~~\hat{e}^{\tilde{\theta}}=e^{-ap^{0}}rd\theta;~~~\hat{e}^{\tilde{\phi}}=e^{-ap^{0}}rsin\theta d\phi.
\ee
Using the relation connecting the old and new coordinate system, $\hat{e}^{a}=\hat{e}_{\mu}^{~a}d\hat{x}^{\mu}$, we find the tetrads as,
\be \label{tetrad}
\hat{e}_{t}^{~\tilde{t}}=e^{\Phi};~~~\hat{e}_{r}^{~\tilde{r}}=\frac{1}{\sqrt{1-\frac{b(r)}{r}}};~~~\hat{e}_{\theta}^{~\tilde{\theta}}=r;~~~\hat{e}_{\phi}^{~\tilde{\phi}}=rsin\theta.
\ee
Now to find the inverse tetrads, we use the equation $\hat{e}_{a}^{~\mu}=\hat{g}^{\mu\nu}\hat{\eta}_{ab}\hat{e}_{\nu}^{~b}$, which gives,
\be \label{inversetetrad}
\hat{e}_{\tilde{t}}^{~t}=e^{-\Phi};~~~\hat{e}_{\tilde{r}}^{~r}=e^{2ap^{0}}\sqrt{1-\frac{b(r)}{r}};~~~\hat{e}_{\tilde{\theta}}^{~\theta}=e^{2ap^{0}}\frac{1}{r};~~~\hat{e}_{\tilde{\phi}}^{~\phi}=e^{2ap^{0}}\frac{1}{rsin\theta}.
\ee
Note that the tetrads (eq.(\ref{tetrad}), eq.(\ref{inversetetrad})) satisfy the relation, $\hat{e}_{a}^{~\mu}\hat{e}_{\nu}^{~a}=\hat{\delta}_{\nu}^{\mu}$ where $\hat{\delta}_{0}^{0}=\delta_{0}^{0}$, $\hat{\delta}_{i}^{j}=\delta_{i}^{j}~e^{2ap^{0}}$.

The Reimann tensor ($\hat{R}^{\rho}_{~\sigma\mu\nu}$) obtained from eq.(\ref{deformedwormhole}) is transformed to the Reimann tensor in the proper frame ($\hat{R}^{a}_{~bcd}$) using $\hat{R}^{a}_{~bcd}=\hat{e}^{a}_{~\rho}\hat{e}^{\sigma}_{~b}\hat{e}^{\mu}_{~c}\hat{e}^{\nu}_{~d}\hat{R}^{\rho}_{~\sigma\mu\nu}$. Using these we find the deformed Einstein tensor in the proper reference frame whose components are,
\bea \label{einsteintensor}
\hat{G}_{\tilde{t}\tilde{t}}&=&e^{8ap^{0}}\frac{b^{\prime}}{r^2},\\ \nonumber
\hat{G}_{\tilde{r}\tilde{r}}&=&e^{4ap^{0}}\left(1-\frac{b}{r}\right)\frac{2\Phi^{\prime}}{r}-e^{6ap^{0}}\frac{b}{r^3},\\ \nonumber
\hat{G}_{\tilde{\theta}\tilde{\theta}}&=&e^{4ap^{0}}\left(1-\frac{b}{r}\right)\left[\Phi^{\prime\prime}+(\Phi^{\prime})^2-\frac{(b^{\prime}r-b)\Phi^{\prime}}{2r^2\left(1-\frac{b}{r}\right)}+\frac{\Phi^{\prime}}{r}\right]+e^{6ap^{0}}\frac{(b-b^{\prime}r)}{2r^3}=\hat{G}_{\tilde{\phi}\tilde{\phi}}.
\eea  
To set up the Einstein field equation, we next construct the deformed energy-momentum tensor.
The energy-momentum tensor is generalized to $\kappa$-deformed space-time using the deformed free particle dispersion relation \cite{neutronstar},
\be \label{deformedEM}
\hat{T}^{\alpha\beta}=\varphi_{\sigma}^{\alpha}\varphi_{\gamma}^{\beta}g(E,a)e^{3ap^{0}}T^{\sigma\gamma}.
\ee
Using the commutative energy-momentum tensor \cite{kipthorne}, we find components of the energy momentum tensor in $\kappa$-deformed space-time to be,
\be \label{EMcomponents}
\hat{T}_{\tilde{t}\tilde{t}}=e^{3ap^{0}}g(E,a)\rho c^2;~~~
\hat{T}_{\tilde{r}\tilde{r}}=-e^{ap^{0}}g(E,a)\tau(r);~~~
\hat{T}_{\tilde{\theta}\tilde{\theta}}=e^{ap^{0}}g(E,a)p(r)=\hat{T}_{\tilde{\phi}\tilde{\phi}}.
\ee
where $\rho c^2$ is the energy density, $\tau(r)$ represents the tension across the unit area measured in the radial direction and $p(r)$ gives the lateral pressure, i.e., pressure measured in directions which are orthogonal to the radial direction. Thus the deformed field equations are,
\bea \label{fieldeq}
b^{\prime}&=&\frac{8\pi G}{c^2}e^{-5ap^{0}}g(E,a)\rho r^2,\\ \nonumber
\Phi^{\prime}&=&\frac{1}{2r(r-b)}\left(e^{2ap^{0}}b-\frac{8\pi G}{c^4}e^{-3ap^{0}}g(E,a)\tau(r)r^3\right),\\ \nonumber
\tau^{\prime}&=&\Phi^{\prime}(\rho c^2-\tau)-\frac{2}{r}(\tau+p)+2ap^{0}\left[\Phi^{\prime}\rho c^2\left(\frac{2\Phi^{\prime}(r-b)}{b^{\prime}}-1\right)+\tau \Phi^{\prime}\right].
\eea
Here $\prime$ stands for derivative with respect to $r$. Note that in the limit $a\rightarrow 0$, all the deformed quantities above reduce to the well known commutative limits. Inverting these equations allows us to change the metric variables $\Phi(r)$ and $b(r)$ as deemed suitable inorder to make the wormhole traversible. This gives us conditions on the stress-energy tensor. In accordance with this requirement, the deformed field equations are written as,
\bea  \label{invertedfieldeq}
\rho &=&\frac{b^{\prime}e^{5ap^{0}}}{\frac{8\pi G}{c^2}g(E,a)r^2},\\ \nonumber
\tau &=& \frac{e^{5ap^{0}}\frac{b}{r}-e^{3ap^{0}}2\Phi^{\prime}(r-b)}{\frac{8\pi G}{c^4}g(E,a)r^2},\\ \nonumber
p&=&\frac{r}{2}[(\rho c^2-\tau)\Phi^{\prime}-\tau^{\prime}]-\tau+ap^{0}[\Phi^{\prime}\rho c^2 r\left(\frac{2\Phi^{\prime}(r-b)}{b^{\prime}}-1\right)+\tau \Phi^{\prime}r].
\eea

\subsection{Spatial geometry of the deformed wormhole}

Consider a $3$-dimensional space having cylindrical symmetry where the distance between any two points on its surface obeys,
\be  \label{euclideansurface}
d\hat{s}^2=e^{-4ap^{0}}\left[dr^{2}\left(1+\left(\frac{dz}{dr}\right)^2\right)+r^{2}d\phi^{2}\right].
\ee
We equate this with the distance relation in the wormhole space-time defined by eq.(\ref{deformedwormhole}) with time fixed to be a constant and $\theta$ set to $\frac{\pi}{2}$, i.e.,
\be \label{wormholesurface}
d\hat{s}^2=e^{-4ap^{0}}\left[\frac{dr^{2}}{\left(1-\frac{b(r)}{r}\right)}+r^{2}d\phi^{2}\right],
\ee
which gives 
\be \label{embeddingfunction}
\frac{dz}{dr}=\pm\left[\frac{r}{b(r)}-1\right]^{-\frac{1}{2}}.
\ee
The solution of the above equation, $z=z(r)$ gives the embedded surface of the wormhole and it is evident that the shape function $b(r)$ dictates the spatial geometry of the wormhole. The above equation is devoid of NC corrections which implies that the non-commutativity of space-time does not affect the spatial geometry of the wormhole. Note that when $r=b(r)=b_{0}$ the above equation diverges, thus the minimum distance $r=b_{0}=b(r)$ is known as the throat radius of the wormhole. Note that eq.(\ref{embeddingfunction}) has no NC corrections as long as $b(r)$ does not have any NC corrections.

To discuss the behaviour of space-time near the throat, one uses proper radial distance defined by,
\be \label{properradialdistance}
\hat{l}(r)=\pm e^{-2ap^{0}}\int \frac{dr}{\sqrt{1-\frac{b(r)}{r}}}.
\ee
One further demands that the proper radial distance has a finite value at all points on the space-time and this requirement on eq.(\ref{properradialdistance}) sets,
\be \label{lcondition}
1-\frac{b(r)}{r} \geq 0.
\ee
It is clear that as $\hat{l}\rightarrow \pm \infty$, $\frac{b(r)}{r} \rightarrow 0$. One also find,
\be \label{extradefinitions}
\frac{dz}{d\hat{l}}=\pm e^{2ap^{0}}\sqrt{\frac{b(r)}{r}};~~~~~~~~\frac{dr}{d\hat{l}}= \pm e^{2ap^{0}}\sqrt{1-\frac{b(r)}{r}}.
\ee

\section{Conditions for traversibility of the $\kappa$-deformed wormhole}

In this section we analyse how the non-commutativity of space-time modifies the traversibility criteria. Since absence of horizon is the primary requirement, we see from the wormhole metric in eq.(\ref{deformedwormhole}) that $\Phi(r)$ necessarily be finite everywhere. Other conditions such as travel time, the force experienced by the traveller are also affected by the non-commutativity of space-time.

\subsection{Time to traverse the wormhole}
Consider a traveller at rest starts to move from a region in the lower universe at $\hat{l}=-\hat{l}_{1}$  through the wormhole to a region in the upper universe at $\hat{l}=\hat{l}_{2}$. The velocity of the traveller as measured by an observer at rest, $\hat{v}(r)$ is given by,
\be  \label{observervelocity}
\hat{v} = \frac{d\hat{l}}{d\hat{\tau}} = g(E,a)\frac{d\hat{l}}{e^{\Phi}dt}= \mp g(E,a)e^{-2ap^{0}}\frac{dr}{\sqrt{1-\frac{b(r)}{r}}e^{\Phi}dt},
\ee
where eq.(\ref{properradialdistance}) is used. We also find,
\be \label{travellervelocity}
\hat{v}\hat{\gamma}=\frac{d\hat{l}}{d{\hat{\tau}}_{T}}= \mp \frac{e^{-2ap^{0}}g(E,a)}{\sqrt{1-\frac{b(r)}{r}}}\frac{dr}{d\tau_{T}}.
\ee
Here, $dt$ is the time measured in $(ct, r, \theta, \phi)$ coordinate (coordinate time lapse), $d\tau$ is the time measured in $(c\tilde{t}, \tilde{r}, \tilde{\theta}, \tilde{\phi})$ (proper time lapse) and $d\tau_{T}$ is the time measured by the traveller in $(c\tilde{t}, \tilde{r}, \tilde{\theta}, \tilde{\phi})$ frame. The $\mp$ sign indicates that the traveller is travelling from the lower universe to the upper universe. Note here $\hat{\gamma}=\frac{1}{\sqrt{1-\frac{\hat{v}^2}{c^2}}}=\frac{1}{\sqrt{1-\hat{\beta}^{2}}}$.

It is reasonable to demand that the duration of the travel should not exceed a year as measured by both the traveller and the observer. This gives us further conditions which are,
\be \label{timemeasured}
\Delta\tau_{T}=\int_{-l_{1}}^{l_{2}}\frac{d\hat{l}}{\hat{v}\hat{\gamma}} \leq 1yr;
~~~~\Delta\hat{t}=\int_{-l_{1}}^{l_{2}}\frac{d\hat{l}}{\hat{v}e^{\Phi}} \leq 1yr.
\ee

\subsection{Acceleration felt by the traveller}

One of the constraints to ensure traversibility through the wormhole is that the acceleration of the traveller should not be greater than the acceleration felt by a freely falling body on earth. Inorder to find the acceleration of the traveller, we need an orthonormal set of basis vectors in the traveller's own reference frame represented by $\hat{e}_{\tilde{t}^{\prime}}, \hat{e}_{\tilde{r}^{\prime}}, \hat{e}_{\tilde{\theta}^{\prime}}, \hat{e}_{\tilde{\phi}^{\prime}}$. The primed coordinates $(c\tilde{t}^{\prime}, \tilde{r}^{\prime}, \tilde{\theta}^{\prime}, \tilde{\phi}^{\prime})$ denote the traveller's reference frame whose basis vectors are determined using the basis vectors of the static observer, $\hat{e}_{\tilde{t}}, \hat{e}_{\tilde{r}}, \hat{e}_{\tilde{\theta}}, \hat{e}_{\tilde{\phi}}$ and the $\kappa$-deformed Lorentz transformation matrix(see Appendix-B). Note that in the Lorentz transformation matrix the traveller's reference frame is considered to be moving in the negative direction with respect to the static observer's reference frame ($\zeta$ is the boost parameter). Thus the orthonormal basis vectors of the traveller are,
\bea \label{travellersbasisvectors}
\hat{e}_{\tilde{t}^{\prime}}&=&[\hat{\gamma}+ap^{0}(4-\zeta\gamma\beta)]\hat{e}_{\tilde{t}}\mp[\hat{\gamma}\hat{\beta}-ap^{0}(\gamma\beta+\gamma\zeta)]\hat{e}_{\tilde{r}}=\hat{u},\\ \nonumber
\hat{e}_{\tilde{r}^{\prime}}&=&[\hat{\gamma}\hat{\beta}+ap^{0}(\gamma\beta-\gamma\zeta)]\hat{e}_{\tilde{t}}\mp[\hat{\gamma}+ap^{0}(4-\zeta\gamma\beta)]\hat{e}_{\tilde{r}},\\ \nonumber
\hat{e}_{\tilde{\theta}^{\prime}}&=&\hat{e}_{\tilde{\theta}},\\ \nonumber
\hat{e}_{\tilde{\phi}^{\prime}}&=&\hat{e}_{\tilde{\phi}}.
\eea
Note that the $\hat{e}_{\tilde{t}^{\prime}}$ represents the four velocity of the traveller and that the $\mp$ sign in the above equations is due to the fact that the radial motion is from the lower universe($-l_{1}$) to the upper universe ($l_{2}$).

Note that the traveller's acceleration in $\kappa$-deformed space-time is given by,
\be \label{acceleration-1}
\hat{a}^{\tilde{\alpha}^{\prime}}=\hat{u}^{\tilde{\alpha}^{\prime}};_{\tilde{\beta}^{\prime}}\hat{u}^{\tilde{\beta}^{\prime}}c^{2},
\ee
where $;$ implies covariant differentiation.
As in the commutative space-time, the four acceleration is always orthogonal to four velocity and we get $\hat{a}_{\tilde{t}^{\prime}}=0$ and since the traveller moves only along the radial direction, we have $\hat{a}_{\tilde{\theta}^{\prime}}=\hat{a}_{\tilde{\phi}^{\prime}}=0$. Thus we have $\mathbf{\hat{a}}=\hat{a}\hat{e}_{\tilde{r}^{\prime}}$ and we find, 
\bea \label{acceleration-2} \nonumber
\hat{a}_{t}&=&\mathbf{\hat{a}}\cdot \hat{e}_{t} = (\hat{a}\hat{e}_{\tilde{r}^{\prime}})\cdot \hat{e}_{t},\\ \nonumber
&=&\hat{a}\left([\hat{\gamma}\hat{\beta}+ap^{0}(\gamma\beta-\gamma\zeta)]\hat{e}_{\tilde{t}}\mp[\hat{\gamma}+ap^{0}(4-\zeta\gamma\beta)]\hat{e}_{\tilde{r}}]\right)\cdot \hat{e}_{t},\\ \nonumber
&=&\hat{a}\left([\hat{\gamma}\hat{\beta}+ap^{0}(\gamma\beta-\gamma\zeta)]\hat{e}_{\tilde{t}}\right)\cdot\hat{e}_{t},\\ 
&=&-\hat{a}\left[\hat{\gamma}\hat{\beta}+ap^{0}(\gamma\beta-\gamma\zeta)\right]e^{\Phi}.
\eea
In writing the last step, we used $\hat{e}_{\tilde{t}}\cdot\hat{e}_{t}=-e^{\Phi}$.

We also find $\hat{a}_{t}$ from eq.(\ref{acceleration-1}),i.e., $\hat{a}_{t}=c^2(\hat{u}_{t,\beta}-\hat{\Gamma}^{\rho}_{t\beta}\hat{u}_{\rho})\hat{u}^{\beta}$, using the four velocity components in the $(ct, r, \theta, \phi)$ coordinates. The four velocity is obtained using the tetrads in eq.(\ref{tetrad}), eq.(\ref{inversetetrad}) and by raising and lowering the indices using the corresponding metric. Thus we find the components of four velocity in $(ct, r, \theta,\phi)$ as
\bea \label{acceleration-4} \nonumber
\hat{u}^{t}&=&e^{-\Phi}(\hat{\gamma}+ap^{0}(4-\zeta\gamma\beta)),\\ \nonumber
\hat{u}_{t}&=&-e^{\Phi}(\hat{\gamma}+ap^{0}(4-\zeta\gamma\beta)),\\ 
\hat{u}^{r}&=&\mp\sqrt{1-\frac{b}{r}}(\hat{\gamma}\hat{\beta}+ap^{0}(\gamma\beta-\zeta\gamma)),\\ \nonumber
\hat{u}_{r}&=&\mp\frac{1}{\sqrt{1-\frac{b}{r}}}(\hat{\gamma}\hat{\beta}-ap^{0}(3\gamma\beta+\zeta\gamma)).
\eea
Substituting the above in $\hat{a}_{t}=c^2(\hat{u}_{t,\beta}-\hat{\Gamma}^{\rho}_{t\beta}\hat{u}_{\rho})\hat{u}^{\beta}$ with the corresponding Christoffel symbols, we find,
\be \label{acceleration-5} 
\frac{\hat{a}_{t}}{c^2}=\pm \sqrt{1-\frac{b}{r}}\left(\hat{\gamma}\hat{\beta}(e^{\Phi}\hat{\gamma})^{\prime}+ap^{0}\left[\gamma\beta(e^{\Phi}(4-\zeta\gamma\beta))^{\prime}+(\gamma\beta-\gamma\zeta)(e^{\Phi}\gamma)^{\prime}\right]\right),
\ee
where $\prime$ on RHS denotes derivative with respect to $r$. Equating eq.(\ref{acceleration-5}) and eq.(\ref{acceleration-2}), we find,
\be \label{acceleration-6}
\hat{a}=\mp \sqrt{\left(1-\frac{b}{r}\right)} e^{-\Phi} c^2\left[(e^{\Phi}\hat{\gamma})^{\prime}+ap^{0}[(e^{\Phi}(4-\zeta\gamma\beta))^{\prime}]\right].
\ee
Traversibility of the wormhole demands that the above acceleration experienced by the traveller should not exceed $1~g_{earth}$. Imposing this and changing all the derivatives with respect to the proper radial coordinate $l$, we find the constraint in $\kappa$-deformed pace-time as,
\be \label{accelerationconstraint} 
\abs{e^{-\Phi}\left[\frac{d}{dl}(e^{\Phi}\hat{\gamma})+ap^{0}\left(\frac{d}{dl}(e^{\Phi}(4-\zeta\gamma\beta))\right)\right]}  \leq \frac{g_{earth}}{c^2}.
\ee

\subsection{Tidal acceleration felt by the traveller}

Another constraint that ensures traversibility of the wormhole is that the tidal acceleration endured by the traveller is of the order of $g_{earth}$. The tidal acceleration between two parts of the traveller's body is given by,
\be \label{tidalacc-1}
\Delta\hat{a}^{\tilde{\alpha}^{\prime}}=-c^2\hat{R}^{\tilde{\alpha}^{\prime}}_{~\tilde{\beta}^{\prime}\tilde{\gamma}^{\prime}\tilde{\delta}^{\prime}}\hat{u}^{\tilde{\beta}^{\prime}}\hat{u}^{\tilde{\gamma}^{\prime}}\hat{\xi}^{\tilde{\delta}^{\prime}}.
\ee
Since the distance vector between these two points is purely spatial, $\hat{\xi}^{\tilde{t}^{\prime}}=0$ and $\hat{u}^{\alpha^{\prime}}=\delta_{\tilde{t}^{\prime}}^{\alpha^{\prime}}$, and thus,
\be \label{tidalacc-2}
\Delta\hat{a}^{\tilde{j}^{\prime}}=-c^2\hat{R}^{\tilde{j}^{\prime}}_{~\tilde{0}^{\prime}\tilde{k}^{\prime}\tilde{0}^{\prime}}\hat{u}^{\tilde{0}^{\prime}}\hat{u}^{\tilde{0}^{\prime}}\hat{\xi}^{\tilde{k}^{\prime}}=-c^2\hat{\eta}^{\tilde{l}^{\prime}\tilde{j}^{\prime}}\hat{R}_{\tilde{l}^{\prime}\tilde{0}^{\prime}\tilde{k}^{\prime}\tilde{0}^{\prime}}\hat{\xi}^{\tilde{k}^{\prime}}.
\ee
We transform the Reimann tensor from the observer's frame to the traveller's frame using the Lorentz transformation and find each of the components of the tidal acceleration in $\kappa$-deformed space-time to be,
\be \label{tidalacc-3}
\Delta\hat{a}^{\tilde{1}^{\prime}}=c^2\left(1-\frac{b}{r}\right)\left[-\Phi^{\prime\prime}+\frac{(b^{\prime}r-b)\Phi^{\prime}}{2r^2\left(1-\frac{b}{r}\right)}-(\Phi^{\prime})^2\right]\left[1+2ap^{0}(1+8\gamma^{3}(1-\beta^2))\right]\hat{\xi}^{\tilde{1}^{\prime}}.
\ee
\begin{multline} \label{tidalacc-4}
\Delta\hat{a}^{\tilde{2}^{\prime}}=-(1+2ap^{0})c^2\hat{\xi}^{\tilde{2}^{\prime}}\left(\frac{\hat{\gamma}^2}{2r^2}\left[2\Phi^{\prime}(r-b)+\hat{\beta}^{2}\left(b^{\prime}-\frac{b}{r}\right)\right]\right)\\
-2ap^{0}c^2\hat{\xi}^{\tilde{2}^{\prime}}\left(\frac{\gamma}{r^2}\left[(4-\zeta\gamma\beta)(r-b)\Phi^{\prime}-\frac{\beta}{2r}(\gamma\beta+\gamma\zeta)(b^{\prime}r-b)\right]\right).
\end{multline}
Note that $\Delta\hat{a}^{\tilde{2}^{\prime}}=\Delta\hat{a}^{\tilde{3}^{\prime}}$ and in the limit $a \rightarrow 0$, we get the commutative result from the above. Now we impose the constraint that the tidal acceleration experienced by the traveller should not be more than $1 g_{earth}$ for a traveller of height $2m$, i.e., $\abs{\hat{\xi}}\approx 2m$. This gives us two more constraints, namely the radial tidal constraint and the lateral tidal constraint in $\kappa$-deformed space-time which respectively are,
\be \label{radialtidalconstraint}
\abs{\left(1-\frac{b}{r}\right)\left[-\Phi^{\prime\prime}+\frac{(b^{\prime}r-b)\Phi^{\prime}}{2r^2\left(1-\frac{b}{r}\right)}-(\Phi^{\prime})^2\right]\left[1+2ap^{0}(1+8\gamma^{3}(1-\beta^2))\right]} \leq \frac{g_{earth}}{c^2~2m},
\ee
\begin{multline} \label{lateraltidalconstraint}
\left| (1+2ap^{0})\frac{\hat{\gamma}^{2}}{2r^2}\left[2\Phi^{\prime}(r-b)+\hat{\beta}^{2}\left(b^{\prime}-\frac{b}{r}\right)\right]+2ap^{0}\frac{\gamma}{r^2}\left[\Phi^{\prime}(4-\zeta\gamma\beta)(r-b)-\frac{\beta}{2r}(\gamma\beta+\gamma\zeta)(b^{\prime}r-b)\right]\right|\\
 \leq \frac{g_{earth}}{c^2~2m}.
\end{multline}
The radial tidal constraint (eq.(\ref{radialtidalconstraint})) can be seen to constrict the redshift function $\Phi$ and the lateral tidal constraint is imposing constraint on the velocity of the traveller \cite{kipthorne}.

Taking the NC correction to tidal acceleration to be within the error bar in the measurement of $g_{earth}$ we find,
\be \label{RTC-1}
\abs{\left(1-\frac{b}{r}\right)\left[-\Phi^{\prime\prime}+\frac{(b^{\prime}r-b)\Phi^{\prime}}{2r^2\left(1-\frac{b}{r}\right)}-(\Phi^{\prime})^2\right]\left[2ap^{0}(1+8\gamma^{3}(1-\beta^2))\right]} \approx \frac{\Delta g_{earth}}{c^2~2m},
\ee
where $\Delta g_{earth}=1.5\times 10^{-5}m/s^{2}$ \cite{clarke}. We know that the commutative part satisfies \cite{kipthorne},
\be \label{RTC-2}
\left(1-\frac{b}{r}\right)\left[-\Phi^{\prime\prime}+\frac{(b^{\prime}r-b)\Phi^{\prime}}{2r^2\left(1-\frac{b}{r}\right)}-(\Phi^{\prime})^2\right] \approx \frac{g_{earth}}{c^2~2m}.
\ee
Substituting eq.(\ref{RTC-2}) in eq.(\ref{RTC-1}), we find an upper bound on $ap^{0}$ as $10^{-8}$ for which the velocity of the traveller is positive with the magnitude of $2.9831\times 10^{8}m/s$. Now in the S.I. unit system $ap^{0} \approx \frac{ap^{0}}{\hbar}=\frac{aE}{c\hbar}=10^{-8}$, and if we take $E=10^{25}eV$ (which is the GUT energy scale), we get the value of deformation parameter $a$ to be $1.9746\times10^{-40}m$ and for energy being $E=1.9746\times10^{20}eV$ we get the deformation parameter as $a=10^{-35}m$ which is the Planck length. 

Using the upper bound on $ap^{0}$ obtained above, we analyse the lateral tidal constraint (eq.(\ref{lateraltidalconstraint})). Using the commutative result \cite{kipthorne} and taking the NC correction to be within the error bar in the measurement of $g_{earth}$, we find that 
\be \label{additionalLTC}
\left[\frac{\gamma}{r^2}(4-\zeta\gamma\beta)(r-b)\Phi^{\prime}-\frac{\beta}{2r^3}(\gamma\beta+\gamma\zeta)(b^{\prime}r-b)\right]\leq 0.0414\times10^{-13}m^{-2}.
\ee
Analysis of the lateral tidal constraint with $\Phi=\Phi_{0}$ gives no additional conditions close to the throat ($b_{0}=r=b$). 

Using the upper bound obtained on $ap^{0}$, one can also analyse the constraint on the acceleration felt by the traveller (eq.(\ref{accelerationconstraint})). Similar to the above analysis, the NC correction is taken to be the error bar in the measurement of $g_{earth}$ which gives an additional constraint,
\be \label{extraaccelerationconstraint}
\left(e^{-\Phi}\frac{d}{dl}(e^{\Phi}(4-\zeta\gamma\beta))\right) \leq 0.2\times10^{-13}
m^{-1}.
\ee
  
\section{Compatibilty conditions on the stress-energy tensor and the wormhole in $\kappa$-deformed space-time}

In previous sections we saw that the conditions for ensuring traversibility of the wormhole put constraints on the components of the wormhole metric. These in turn set constraints on the stress-energy tensor as well, via the field equations. To analyse the behaviour of the tension close to the throat \cite{kipthorne}, a dimensionless parameter $\hat{\sigma}$ is defined as,
\be \label{zeta-1}
\hat{\sigma}=\frac{\hat{\tau}-\hat{\rho}c^2}{\lvert\hat{\rho}c^2\rvert}=\frac{e^{-2ap^{0}}\tau-\rho c^2}{\lvert \rho c^2\rvert},
\ee
where the $\kappa$-deformed stress-energy tensor (eq.(\ref{EMcomponents})) is used. Using the field equations(eq.(\ref{invertedfieldeq})), for $\rho$ and $\tau$, one obtains,
\be \label{zeta-2}
\hat{\sigma}=\frac{\frac{b}{r}-2\Phi^{\prime}(r-b)-b^{\prime}+2ap^{0}\left(4\Phi^{\prime}(r-b)-\frac{b}{r}\right)}{b^{\prime}}.
\ee
A geometrical requirement for the wormhole is that the embedding surface should funnel outwards and be connected to flat space-time as $\hat{l} \rightarrow \pm\infty$. Imposing this flaring out condition,
\be  \label{flaringoutcondition}
\frac{d^2r}{dz^2}=\frac{b-b^{\prime}r}{2b^2} \geq 0,
\ee
in eq.(\ref{zeta-2}),we get,
\be  \label{zeta-3}
\hat{\sigma}=\frac{2b^2}{\abs{b^{\prime}}r}\frac{d^2r}{dz^2}-\frac{2\Phi^{\prime}}{\abs{b^{\prime}}}(r-b)+2ap^{0}\left[\frac{4\Phi^{\prime}}{\abs{b^{\prime}}}(r-b)-\frac{b}{\abs{b^{\prime}}r}\right].
\ee
At or near the throat where $r=b=b_{0}$, eq.(\ref{zeta-3}) becomes,
\be \label{throatzeta}
\hat{\sigma_{0}}=2b_{0}\frac{d^2r}{dz^2}-2ap^{0}.
\ee
Since $\frac{d^2r}{dz^2}\geq 0$, the condition $\hat{\sigma_{0}}\geq 0$ using eq.(\ref{zeta-1}) implies (noting $ap^{0}$ is of the order of $10^{-8}$) that, at or near the throat,
\be \label{throatzeta-2}
\hat{\sigma_{0}}=\frac{\hat{\tau}_{0}-\hat{\rho}_{0}c^2}{\lvert \hat{\rho}_{0}c^{2}\rvert}\geq 0
\ee
which implies that $\hat{\tau}_{0}-\hat{\rho}_{0}c^2 \geq 0$,i.e., tension at the throat of the material generating the wormhole should exceed its energy density. This is the defining property of exotic material; thus the necessity of exotic material for the generation of wormhole throat persists in $\kappa$-deformed space-time as in the commutative space-time, albeit the quantities get NC corrections.

Inorder to study the implication of exotic materials in $\kappa$-deformed space-time, consider a traveller with velocity close to the speed of light moving through the wormhole. The energy density seen by the traveller is,
\be \label{t00}
\hat{T}_{\tilde{0}^{\prime}\tilde{0}^{\prime}}=\hat{\gamma}^{2}g(E,a)\left[\rho c^2-\hat{\beta}^{2}\tau(r)\right]+ap^{0}\left(3\gamma^{2}\rho c^2+2\gamma\rho c^2(4-\zeta\gamma\beta)-\gamma^{2}\beta^{2}\tau+2\gamma\beta\tau(\gamma\beta+\gamma\zeta)\right),
\ee 
which at the throat (with $\tau=\tau_{0}$, $\rho=\rho_{0}$, $v \approx c$) becomes, 
\be \label{throat t00}
\hat{T}_{\tilde{0}^{\prime}\tilde{0}^{\prime}}=g(E,a)\left[\hat{\gamma}^{2}(\rho_{0} c^2-\tau_{0})+\tau_{0}\right]+ap^{0}\left(\gamma^{2}(3\rho_{0} c^2-\tau_{0})+2\gamma\left[\rho_{0} c^2(4-\zeta\gamma)+\tau_{0}\gamma(1+\zeta)\right]\right)
\ee
For the traveller moving sufficiently fast ($\gamma$ is large), the person will see a negative energy density when $\tau_{0} > \rho_{0}c^2$, i.e., the material at the throat is exotic in nature. This condition leads to another constraint due to non-commutativity of space-time which is given by,
\be 
\abs{ \gamma (3\rho_{0}c^2-\tau_{0})} > \abs{ 2\left[\rho_{0}c^2(4-\zeta\gamma)+\tau_{0}\gamma(1+\zeta)\right]},
\ee
if $(3\rho_{0}c^2-\tau_{0}) < 0$.

Since it is undesirable to have exotic matter spread all over the space-time, one approach taken is to confine the exotic matter to a small region around the throat and connecting it to non-exotic matter beyond it (see Appendix-D for details). In this method, Casimir effect has been considered as a possible candidate to produce the necessary exoticity for the wormhole \cite{kipthorne}.

\section{Casimir wormholes in $\kappa$-deformed space-time}

Casimir effect is the phenomenon where two plane, parallel conducting plates experiences an attractive force due to vacuum fluctuations of the electromagnetic field when the separation distance between them is of the order of a few microns. This effect was first predicted by H. G. B. Casimir in 1948 and has hence become experimentally verified \cite{lambrecht}. Casimir effect is of interest in the study of wormhole solutions due to the negative nature of Casimir energy density and the violation of null energy condition by the Casimir force \cite{sahoo}. Violation of null energy condition is exhibited by exotic materials which is necessary for the existence of traversible wormholes. The energy density due to Casimir effect between two parallel, conducting plates is,
\be \label{casimirenergy}
\rho(L)=-\frac{\pi^{2}\hbar c}{720 L^{4}},
\ee 
where $L$ gives the separation between the plates. The plate separation $L$ is replaced by the radial coordinate $r$. This energy density is substituted in the first of the deformed field equation(eq.(\ref{fieldeq})) which gives a solution for the shape function, $b(r)$, as,
\be \label{b-1}
b(r)=e^{-5ap^{0}}g(E,a)\frac{8\pi G}{c^4}\frac{\pi^{2}\hbar c}{720}\frac{1}{r}+C,
\ee
where $C$ is the integration constant which is fixed by imposing the throat condition $b(r)=r_{0}$ at $r=r_{0}$. This gives,
\be \label{b-2}
b(r)=r_{0}+(1+ap^{0}\epsilon)\alpha\left[\frac{1}{r}-\frac{1}{r_{0}}\right],
\ee
where
\be 
\epsilon = \left(\frac{1}{2}\left(1-\left(\frac{mc^2}{E}\right)^2\right)-5\right);~~~~
\alpha=\frac{8\pi G}{c^3}\frac{\pi^{2}\hbar }{720}=\frac{\pi^{3}}{90}l_{p}^{2},
\ee
here $l_{p}$ is the Planck length. Inorder to find the redshift function ($\Phi(r)$) one needs to solve the second field equation(eq.(\ref{fieldeq}))using eq.(\ref{b-2}) and with $-\tau(r)=P_{r}(r)$ where $P_{r}(r)$ is the radial pressure (which is the negative of tension per unit area in the radial direction) and the equation of state $P_{r}(r)=\omega \rho(r)$. Here $\omega$ is the equation of state parameter. This gives,
\begin{multline} \label{redshift-1}
\Phi(r)=\Phi(r_{0})-\frac{1}{2}\left(\left[\frac{(1-\omega)+ap^{0}\left(2+\epsilon-\omega\Delta\right)}{(1+ap^{0}\epsilon)}\right]\ln(r)+\frac{\alpha\omega-r_{0}^{2}-ap^{0}(2r_{0}^{2}-\alpha\omega\Delta)}{(r_{0}^{2}+\alpha+ap^{0}\alpha\epsilon)}\ln(r-r_{0})\right)\\
-\frac{1}{2}\left(\frac{r_{0}^{2}\omega-\alpha-ap^{0}(2\alpha(1+\epsilon)-\omega\Delta r_{0}^{2})}{(1+ap^{0}\epsilon)(r_{0}^{2}+\alpha+ap^{0}\alpha\epsilon)}\ln(rr_{0}+\alpha+ap^{0}\alpha\epsilon)\right),
\end{multline} 
where, 
\be 
\Delta= \left(\frac{1}{2}\left(1-\left(\frac{mc^2}{E}\right)^2\right)-3\right).
\ee
At $r=r_{0}$ in eq.(\ref{redshift-1}), the second term leads to divergences in the metric when $(\alpha\omega-r_{0}^{2}-ap^{0}(2r_{0}^{2}-\alpha\omega\Delta))>0$, leading to the formation of a black hole. Inorder to prevent this, one imposes $(\alpha\omega-r_{0}^{2}-ap^{0}(2r_{0}^{2}-\alpha\omega\Delta))=0$ which gives,
\be \label{omega}
\omega=\frac{r_{0}^{2}}{\alpha}\left(1+ap^{0}(2-\Delta)\right).
\ee
Substituting for $\alpha$ from eq.(\ref{omega}) in eq.(\ref{redshift-1}) gives,
\be \label{redshiftfunction}
\Phi(r)=\frac{1}{2}(\omega-1)(1+2ap^{0})\ln{\left[\frac{\omega r}{\omega r+r_{0}}\right]}.
\ee
Note as $r \rightarrow \infty$, $\Phi(r_{0})=0$ and in the limit $a \rightarrow 0$, we get back the commutative result. Substituting for $\alpha$ and using eq.(\ref{omega}) in eq.(\ref{b-2}), we get the shape function as,
\be \label{shapefunction}
b(r)=r_{0}\left(1-\frac{1}{\omega}\right)+\frac{r_{0}^{2}}{\omega r}.
\ee
Note that the shape function has no NC corrections. Thus the explicit form of $\kappa$-deformed wormhole metric is,
\be \label{casimirwormhole}
d\hat{s}^{2}=-\left[\frac{\omega r}{\omega r+r_{0}}\right]^{(\omega-1)(1+2ap^{0})}c^2dt^{2}+e^{-4ap^{0}}\left[\frac{dr^{2}}{\left(1-\frac{r_{0}}{r}\left(1-\frac{1}{\omega}\right)-\frac{r_{0}^{2}}{\omega r^{2}}\right)}+r^{2}d\theta^{2}+r^2sin^{2}\theta d\phi^{2}\right].
\ee
For $\omega=3$, the above metric becomes,
\be \label{casimirmetric-3}
d\hat{s}^{2}=-\left[\frac{3r}{3r+r_{0}}\right]^{2(1+2ap^{0})}c^2dt^{2}+e^{-4ap^{0}}\left[\frac{dr^{2}}{\left(1-\frac{2r_{0}}{3r}-\frac{r_{0}^{2}}{3r^{2}}\right)}+r^{2}d\theta^{2}+r^2sin^{2}\theta d\phi^{2}\right],
\ee
where the redshift function is
\be \label{redshiftfunction-3}
\Phi(r)=(1+2ap^{0})\ln{\left[\frac{3r}{3r+r_{0}}\right]}.
\ee
and the shape function is
\be \label{shapefunction-3}
b(r)=\frac{2}{3}r_{0}+\frac{r_{0}^{2}}{3r}.
\ee

Next we consider the third field equation obtained by using the formula, $\hat{G}_{\tilde{\theta}\tilde{\theta}}=\frac{8\pi G}{c^4}\hat{T}_{\tilde{\theta}\tilde{\theta}}$ and substitutions are done for $b(r)$ and $\Phi(r)$ using eq.(\ref{shapefunction}) and eq.(\ref{redshiftfunction}), respectively. $\hat{T}_{\tilde{\theta}\tilde{\theta}}$ is the tangential pressure and is calculated using the equation of state $P_{t}(r)=\omega_{t}(r)\rho(r)$. This leads to,
\be 
\omega_{t}(r)=-\frac{\omega^{2}(4r-r_{0})+r_{0}(4\omega +1)+ap^{0}\left[\omega^{2}(r-2r_{0})+\omega(4r_{0}-3r)+2(r-r_{0})\right]}{4(\omega r+r_{0})}.
\ee
In the limit $a \rightarrow 0$, the commutative result is obtained.

The proper radial distance (eq.(\ref{properradialdistance})) for $\kappa$-deformed Casimir wormhole is
\be 
\hat{l}(r)=\pm (1-2ap^{0})\left[\frac{\sqrt{3(3r+r_{0})(r-r_{0})}}{3}+\frac{r_{0}}{3}\ln{\left(\frac{3r-r_{0}+\sqrt{3(3r+r_{0})(r-r_{0})}}{2r_{0}}\right)}\right],
\ee
and the proper travel time (eq.(\ref{timemeasured})) for a traveller moving through the Casimir wormhole in $\kappa$-deformed space-time with the velocity $v(r)$ is 
\be \label{traveltime}
\Delta\hat{\tau}_{T}=\frac{(1-2ap^{0})g(E,a)}{v\gamma}\left[\frac{\sqrt{3(3r+r_{0})(r-r_{0})}}{3}+\frac{r_{0}}{3}\ln{\left(\frac{3r-r_{0}+\sqrt{3(3r+r_{0})(r-r_{0})}}{2r_{0}}\right)}\right].
\ee
Note that the travel time reduces due to NC corrections.

The embedding surface remains same as in the commutative case \cite{garattini} as neither the equation (eq.(\ref{embeddingfunction})) nor the shape function(eq.(\ref{shapefunction-3})) has NC corrections.

The acceleration felt by the traveller while travelling through $\kappa$-deformed Casimir wormhole can be found using eq.(\ref{accelerationconstraint}) and substituting for the redshift function and shape function obtained in eq.(\ref{redshiftfunction-3}) and eq.(\ref{shapefunction-3}), respectively. Also the traveller is assumed to move with constant speed and $\gamma \approx 1$ \cite{garattini}. Thus the acceleration constraint in $\kappa$-deformed space-time reduces to,
\be \label{accelerationconstraintcasimir}
\abs{\hat{a}}=\abs{\sqrt{1-\frac{2r_{0}}{3r}-\frac{r_{0}^{2}}{3r^{2}}}\left[\frac{r_{0}}{r(3r+r_{0})}\left(1+ap^{0}(6-\zeta\beta)\right)\right]} \leq \frac{g_{earth}}{c^2}.
\ee 
Close to the throat, i.e., at $r=r_{0}$, the above equation reduces to zero which implies that the traveller has no acceleration close to the throat as in the commutative case. The additional constraint from acceleration felt by the traveller obtained using the upper bound on $ap^{0}$ and error bar in the measurement of $g_{earth}$ (eq.(\ref{extraaccelerationconstraint})) reduces to,
\be \label{extraaccelerationconstraintcasimir}
\frac{1}{\sqrt{3}r(3r+r_{0})}\left(r_{0}(1+2ap^{0})(4-\zeta\beta)\sqrt{3-\frac{r_{0}^{2}}{r^{2}}-\frac{2r_{0}}{r}}\right) \leq 0.2\times 10^{-13}m^{-1}.
\ee

The radial tidal constraint (eq.(\ref{radialtidalconstraint})) for the $\kappa$-deformed Casimir wormhole becomes,
\be \label{radialTCcasimir}
\abs{\frac{r_{0}\left(6r^{2}-7rr_{0}-r_{0}^{2}+2ap^{0}r_{0}(r_{0}-r)\right)}{3r^{4}(3r+r_{0})}\left[1+4ap^{0}(1+4\gamma^{3}(1-\beta^{2}))\right]}\leq (10^{8}m)^{-2}.
\ee
Close to the throat where $r\approx r_{0}$, we get,
\be \label{throatradius}
r_{0}\left[1-2ap^{0}(1+4\gamma^{3}(1-\beta^{2}))\right] \geq 10^{8}m.
\ee
Note that the the lower bound on the throat radius increases due to $\kappa$-deformed space-time.

The lateral tidal constraint in $\kappa$-deformed space-time (eq.(\ref{lateraltidalconstraint})) close to the throat reduces to,
\be \label{lateraltidalconstraintcasimir}
\abs{-2v^{2}+4ap^{0}v\zeta c}\leq \frac{3g_{earth}r_{0}^{2}}{2},
\ee
physically acceptable solution of the above equation gives
\be \label{velocity} 
v\leq \frac{1}{2}\left(2ap^{0}\zeta c + \sqrt{3r_{0}^{2}g_{earth}}\right).
\ee
We see that the upper bound on the velocity of the traveller increases due to non-commutativity.

The additional constraint obtained using the upper bound on $ap^{0}$ and error bar in the measurement of $g_{earth}$ on the lateral tidal constraint(eq.(\ref{additionalLTC})) close to the throat($r \approx r_{0}$) reduces to,
\be \label{additional LTC casimir} 
\frac{2\beta(\beta+\zeta)}{3r_{0}^{2}}\leq 0.0414\times 10^{-13}m^{-2}.
\ee

The volume integral quantifier gives information about the total amount of exotic matter in the space-time \cite{visserIV} and is given in $\kappa$-deformed space-time as,
\bea \label{volumeintegralquantifier}
I_{v}&=&\int\left(\rho(r)+P_{r}(r)\right)dV\\ \nonumber
&=&-\frac{4r_{0}}{3\left(\frac{8\pi G}{c^4}\right)}\left[1-ap^{0}\left(\frac{1}{2}\left(1-\left(\frac{mc^{2}}{E}\right)^{2}\right)-5\right)\right]\\ \nonumber
&=&-\frac{4r_{0}}{3\left(\frac{8\pi G}{c^4}\right)}\left[1-ap^{0}\epsilon\right].
\eea
Here we see that the amount of exotic matter necessary for wormhole's curvature decreases due to non-commutativity of space-time.

From eq.(\ref{omega}) for $\omega=3$, we have,
\be \label{casimirwormholesize}
r_{0}=\sqrt{3\alpha}(1+\frac{ap^{0}}{2}\epsilon)=\sqrt{\frac{3\pi^{3}}{90}}l_{p}(1+\frac{ap^{0}}{2}\epsilon).
\ee
This implies that $\kappa$-deformed traversible wormhole solution obtained from Casimir energy density has size of the order of Planck length. This result complies with the commutative results \cite{garattini}. From the radial tidal constraint in $\kappa$-deformed space-time (eq.(\ref{radialTCcasimir})), we find that,
\be 
\abs{\xi}\leq 10^{-86}\left(1+ap^{0}(\epsilon+4(1+4\gamma^{3}(1-\beta^{2})))\right)m,
\ee
which means humans cannot traverse through this wormhole as in the commutative space-time.

\section{Conclusion}

One of the solutions of General Relativity is that of the wormhole which is a tunnel that connects two different regions of the universe. The wormhole solution is constructed and analysed in $\kappa$-deformed space-time in this paper. The deformed traversible wormhole metric is constructed from which field equations are calculated in the proper reference frame using tetrads. It was seen that the spatial geometry of the wormhole is unaffected by the non-commutativity of space-time as the shape function, $b(r)$, has no NC corrections. Further conditions that ensure traversibility of the wormhole are analysed in $\kappa$-deformed space-time. On analysing the radial tidal constraint in $\kappa$-deformed space-time, an upper bound of $10^{-8}$ is obtained on $ap^{0}$ for which the traveller's velocity is $2.9831\times 10^{8}m/s$. For this obtained upper bound, the deformation parameter, $a$, takes the value of $1.9746\times 10^{-40}m$ for energy being $10^{25}eV$ and the deformation parameter becomes $10^{-35}m$, when the energy of the particle probing the space-time is $1.9746\times 10^{20}eV$. The acceleration constraint and the lateral tidal constraint lead to additional constraints in $\kappa$-deformed space-time on substituting this upper bound on $ap^{0}$ and taking the NC corrections to be within the error bar in the measurement of $g_{earth}$. All these conditions that ensure traversibility of the wormhole impose constraints on the metric variables which in turn puts constraints on the stress-energy tensor through the deformed field equations. Analysis of these constrains shows that exotic matter is required for the generation of wormhole geometry in $\kappa$-deformed space-time as in the commutative space-time. It is also shown that a traveller moving through the deformed wormhole with velocity close to the speed of light will see a negative energy density for the matter threading the wormhole.

Next, we look into the effect of using Casimir energy as a source for exoticity on the deformed traversible wormhole. The metric variables $b(r)$ and $\Phi(r)$ are solved using the equation of state and Casimir energy as the energy density in the stress-energy tensor. The solution for $\Phi(r)$ leads to the creation of a black hole. To avoid this, a condition is imposed on $\omega$ (eq.(\ref{omega})) and the explicit form of the deformed wormhole metric is obtained. Further, the constraints for a traversible wormhole is analysed for Casimir energy and the lower bound on the throat radius increases due to $\kappa$-deformed space-time. The travel time as measured by the traveller was found to decrease due to $\kappa$-deformed space-time. Further, velocity of the traveller traversing the deformed Casimir wormhole is also determined whose upper bound is found to increase because of non-commutativity. The amount of exotic matter determined by the volume integral quantifier decreases in $\kappa$-deformed space-time compared to the commutative case.

\section{Acknowledgement}

HS thanks Prime Minister Research Fellowship (PMRF id:3703690) for the financial support.

\renewcommand{\thesection}{Appendix-A}
\section{$\kappa$-deformed Minkowski metric}
\renewcommand{\thesection}{A}

We have \cite{sir, zuhair}; 
\be   \label{metricommutation}
[\hat{x}_{\mu},\hat{P}_{\nu}]=i\hat{\eta}_{\mu \nu}, 
\ee
where $\hat{x}_{\mu}=x_{\alpha}\varphi_{\mu}^{\alpha}$ and $\hat{P}_{\nu}=\eta_{\alpha\beta}(\hat{y})p^{\beta}\varphi_{\nu}^{\alpha}$. Substituting these into the commutation relation,
\be \label{deformedminkowskimetric}
 \hat{\eta}_{\mu\nu}=\eta_{\alpha\beta}(\hat{y})\Big(p^{\beta}\frac{\partial \varphi^{\alpha}_{\nu}}{\partial p^{\sigma}}\varphi_{\mu}^{\sigma}+\varphi_{\mu}^{\beta}\varphi_{\nu}^{\alpha}\Big),
\ee
where $\hat{y}_{0}=x_{0}-ax_{j}p^{j}$ and $\hat{y}_{i}=x_{i}$. Using the realization $\varphi_{0}^{0}=1$ and $\varphi_{i}^{j}=\delta_{i}^{j}e^{-ap^{0}}$ gives the elements of the Minskowski metric as,
  \be \label{minkowski metric components}
\begin{aligned}
\hat{\eta}_{00}&=\eta_{00}(\hat{y})=\eta_{00}(x_{i}),\\
\hat{\eta}_{0i}&=\eta_{i0}(\hat{y})\big(1-ap^0\big)e^{-ap^0}-ap^me^{-ap^0}\eta_{im}(\hat{y}),\\ 
\hat{\eta}_{i0}&=\eta_{0i}(\hat{y})e^{-ap^0},\\
\hat{\eta}_{ij}&=\eta_{ji}(\hat{y})e^{-2ap^0}.
\end{aligned}
\ee
Thus the Minkowski metric, considering terms only upto first order in $a$, can be written as,
\be \label{deformedminkowski}
\hat{\eta}_{\mu \nu}=\begin{bmatrix}
-1 & -ap^{1} & -ap^{2} & -ap^{3} \\
0 & e^{-2ap^{0}} & 0 & 0 \\
0 & 0 & e^{-2ap^{0}} & 0 \\
0 & 0 & 0 & e^{-2ap^{0}} 
\end{bmatrix}.
\ee

\renewcommand{\thesection}{Appendix-B}
\section{Lorentz Transformation in $\kappa$-deformed space-time}
\renewcommand{\thesection}{B}

Consider the vector $\hat{x}$ which transforms to $\hat{x}^{\prime}=\hat{\Lambda}\hat{x}$. Invariance of $\hat{x}\cdot\hat{x}$ under $\kappa$-deformed Lorentz transformation gives,
\be \label{lt-1}
\hat{\Lambda}^T\hat{\eta}\hat{\Lambda}=\hat{\eta},
\ee
where $\hat{\Lambda}^T$ represents the transpose of the $\kappa$-deformed Lorentz transformation matrix, $\hat{\Lambda}$, and $\hat{\eta}$ is the $\kappa$-deformed Minkowski metric. Considering only proper Lorentz transformation, we define $\hat{\Lambda}=e^{\hat{L}}$, where $\hat{L}$ is a traceless matrix.

By multiplying eq.(\ref{lt-1}) with $\hat{\eta}^{-1}$ from the left and $\hat{\Lambda}^{-1}$ from the right, one finds,
\be  \label{lt-2}
\hat{\eta}^{-1}\hat{\Lambda}^{T}\hat{\eta}=\hat{\Lambda}^{-1}.
\ee
Substituting $\hat{\Lambda}=e^{\hat{L}}$ in the above equation and expanding, one finds
\be \label{lt-3}
\hat{L}^{T}\hat{\eta}=-\hat{\eta}\hat{L},
\ee
where one uses the definition $\hat{\Lambda}^{-1}=e^{-\hat{L}}$. Choose,
\be \label{Lmatrix}
\hat{L}=\begin{bmatrix}
0 & \hat{L}_{01} & \hat{L}_{02} & \hat{L}_{03}\\
\hat{L}_{10} & 0 & \hat{L}_{12} & \hat{L}_{13}\\
\hat{L}_{20} & \hat{L}_{21} & 0 & \hat{L}_{23}\\
\hat{L}_{30} & \hat{L}_{31} & \hat{L}_{32} & 0\\
\end{bmatrix}.
\ee
Substituting the above and the deformed Minkowski metric in eq.(\ref{lt-3}), we find
\bea \label{lt-4}
\hat{L}_{i0}&=&\hat{L}_{0i}(1+2ap^{0})\\ \nonumber
\hat{L}_{ij}&=&-\hat{L}_{ji}.
\eea
for $i,j=1,2,3;~ i\neq j$ and also $p^{1}=p^{2}=p^{3}=0$. Thus one finds the matrix realization of $\kappa$-deformed boost generators as,
\be \label{lt-5}
\hat{K}_{1}=\begin{bmatrix}
0 & 1 & 0 &0\\
1+2ap^{0} & 0 & 0 &0\\
0 & 0 & 0 &0\\
0 & 0 & 0 &0\\
\end{bmatrix};
~~~~\hat{K}_{2}=\begin{bmatrix}
0 & 0 & 1 &0\\
0 & 0 & 0 &0\\
1+2ap^{0} & 0 & 0 &0\\
0 & 0 & 0 &0\\
\end{bmatrix};
~~~~\hat{K}_{3}=\begin{bmatrix}
0 & 0 & 0 &1\\
0 & 0 & 0 &0\\
0 & 0 & 0 &0\\
1+2ap^{0} & 0 & 0 &0\\
\end{bmatrix}.
\ee
Let $\hat{\Lambda}=e^{\hat{L}}=e^{-\vec{\hat{\omega}}\cdot\vec{\hat{\xi}}-\vec{\hat{\zeta}}\cdot\vec{\hat{K}}}$. Choose $\vec{\hat{\omega}}=0$ and $\vec{\hat{\zeta}}=\hat{\zeta}\hat{n}_{1}$ such that $\vec{\hat{\zeta}}\cdot\vec{\hat{K}}=\hat{\zeta}\hat{K}_{1}$. This gives $\hat{\Lambda}=e^{\hat{L}}=e^{-\hat{\zeta}\hat{K}_{1}}$ which on expansion and appropriate simplification leads to the $\kappa$-deformed Lorentz transformation matrix valid upto first order in $a$,
\be \label{lt-6}
\hat{\Lambda}=\begin{bmatrix}
cosh\hat{\zeta}+ap^{0}(4+\zeta sinh\zeta) & -sinh\hat{\zeta}(1-ap^{0})-ap^{0}\zeta cosh\zeta & 0 & 0\\
-sinh\hat{\zeta}(1+ap^{0})-ap^{0}\zeta cosh\zeta & cosh\hat{\zeta}+ap^{0}(4+\zeta sinh\zeta) & 0 & 0\\
0 & 0 & 1& 0\\
0 & 0 & 0 & 1
\end{bmatrix}.
\ee 
where,
\be \label{lt-7}
cosh\hat{\zeta}=\hat{\gamma}=\frac{1}{\sqrt{1-\frac{\hat{v}^2}{c^2}}};~~~~sinh\hat{\zeta}=\hat{\gamma}\hat{\beta}=\frac{\frac{\hat{v}}{c}}{\sqrt{1-\frac{\hat{v}^2}{c^2}}}.
\ee
Substituting this in eq.(\ref{lt-6}), we find the $\kappa$-Lorentz transformation matrix to be,
\be
\hat{\Lambda}=\begin{bmatrix}
\hat{\gamma}+ap^{0}(4+\zeta\gamma\beta) & -\hat{\gamma\beta}+ap^{0}(\gamma\beta-\zeta\gamma) & 0 & 0\\
-\hat{\gamma\beta}-ap^{0}(\gamma\beta+\zeta\gamma) & \hat{\gamma}+ap^{0}(4+\zeta\gamma\beta) & 0 & 0\\
0 & 0 & 1 & 0\\
0 & 0 & 0 & 1
\end{bmatrix}. 
\ee

\renewcommand{\thesection}{Appendix-C}
\section{Boundary Conditions for a wormhole in $\kappa$-deformed space-time}
\renewcommand{\thesection}{C}

As in the commutative case, we take $\rho$ and $p$ to be finite only in the sphere of radies $r=R_{s}$ while $\tau$ goes to zero continuously at $r=R_{s}$. Outside this sphere, one takes geometry to be of the Schwarzschild form. The $\kappa$-deformed Schwarzschild metric is,
\be  \label{deformedschwarzschild}
d\hat{s}^2=-\left[1-\frac{2GM}{c^2 r}\right]c^2dt^2+e^{-4ap^{0}}\left[\frac{dr^{2}}{\left(1-\frac{2GM}{c^2 r}\right)}+r^2d\theta^{2}+r^2sin^{2}\theta d\phi^{2}\right].
\ee 
Comparing the above metric with the deformed wormhole metric (eq.(\ref{deformedwormhole})) at $r=R_{s}$, we find the shape function, $b(r)$, and the redshift function $\Phi(r)$ as,
\bea
b(r) &=&\frac{2GM}{c^2}=b(R_{s})=\text{constant}\\ \nonumber
\Phi(r)&=&\frac{1}{2}\ln\left[1-\frac{B}{r}\right],
\eea
respectively. These solutions are valid in the region $r>R_{s}$.

\renewcommand{\thesection}{Appendix-D}
\section{Wormhole solutions in $\kappa$-deformed space-time when exotic matter is confined in a small region around the throat}
\renewcommand{\thesection}{D}

Inorder to look at wormhole solutions in the vicinity of the throat, one needs to first look at the solutions when the exotic material is confined to a surface radius of $R_{s}$ which is then joined to an exterior Schwarzschild solution. Take,
\be \label{choiceofb-1}
b(r)=(b_{0}r)^{\frac{1}{2}};~~~~\Phi(r)=\Phi_{0}=const.
\ee
Using the above, the first two field equations in eq.(\ref{fieldeq}) become,
\be \label{fieldeqnsforachoice}
\rho =\frac{\frac{1}{2}be^{5ap^{0}}}{\frac{8\pi G}{c^2}g(E,a)r^3};~~~~\tau = \frac{e^{5ap^{0}}b}{\frac{8\pi G}{c^4}g(E,a)r^3}=2\rho c^2.
\ee
The boundary conditions of the wormhole requires $\tau$ to be continuous but allows discontinuities in $p$ and $\rho$ \cite{kipthorne}. A transition layer $\Delta \hat{R}$ is taken which is joined to the exterior Schwarzschild solution at $\hat{R_{s}}+\Delta \hat{R}$. $\tau$ should be brought to zero near the surface radius $R_{s}$ for which one consider the following assumptions for the transition layer in $\kappa$-deformed space-time,
\bea \label{assumptions}
\rho(r)&=&\frac{e^{-2ap^{0}}\tau(R_{s})}{c^2}\left[\frac{R_{s}}{\Delta R}\right], \\ \nonumber
\tau(r)&=&\tau(R_{s})-\frac{\tau(R_{s})}{\Delta R}(r-R_{s}).
\eea
Note that the second assumption has no NC corrections. Substituting for $\rho(r)$ using the above assumption in the first field equation(eq.(\ref{fieldeq})) and integrating to find $b(r)$ gives,
\be \label{transitionlayerb-1}
b(r)=b(R_{s})+e^{-7ap^{0}}\frac{8\pi G}{c^4}g(E,a)\tau(R_{s})\frac{R_{s}}{\Delta R}\frac{(r^{3}-R_{s}^{3})}{3},
\ee
where the integration constant is fixed by finding $b(r)$ at $r=R_{s}$. Using eq.(\ref{fieldeqnsforachoice}) for $\tau(R_{s})$ in the above, we find
\bea  \label{transitionlayerb-2}
b(r)&=&b(R_{s})+e^{-2ap^{0}}\frac{b(R_{s})(r-R_{s})}{\Delta R},\\ \nonumber
B&=&b(R_{s}+\Delta R)=2b(R_{s})(1-ap^{0}).
\eea
Now, using the second field equation(eq.(\ref{fieldeq})), where $b(r)$ is substituted using eq.(\ref{transitionlayerb-2}) and $\tau(r)$ is substituted using the second assumption(eq.(\ref{assumptions})), one finds,
\bea \label{transitionlayerphi}
\Phi^{\prime}(r)&=&\frac{b(R_{s})(r-R_{s})}{R_{s}^{2}\Delta R}\left[1+ap^{0}\left(1-\frac{2(r-R_{s})}{R_{s}}\right)\right],\\ \nonumber
\Phi^{\prime}(R_{s}+\Delta R)&=&\frac{B}{2R_{s}^{2}}(1+ap^{0}).
\eea
In all these calculations, one assumes the thickness of the transition layer to be $\Delta R=b(R_{s})<<R_{s}$. We also have
\bea  \label{transitionlayertau}
\tau(r)&=&\tau(R_{s})-\frac{\tau(R_{s})}{\Delta R}(r-R_{s}), \\ \nonumber
\tau(R_{s}+\Delta R)&=&0.
\eea
and 
\be \label{transitionlayerp}
p(r)=\frac{\tau(R_{s})R_{s}}{2\Delta R}=const;~~~~\rho(r)=\frac{e^{-2ap^{0}}\tau(R_{s})R_{s}}{c^{2}\Delta R}.
\ee
Thus we have the solutions(eq.(\ref{transitionlayerb-2}, \ref{transitionlayerphi}, \ref{transitionlayertau}, \ref{transitionlayerp}) at the transition layer.

For the exotic matter confined to some radius $r_{c}$ beyond which non-exotic matter is present, one chooses
\bea
b(r)&=&(b_{0}r)^{\frac{1}{2}},~~~~~at~b_{0}\leq r \leq r_{c} \\ \nonumber
\Phi&=&\Phi_{0}.
\eea
The size of the wormhole and the travel time is determined by the slope of the embedding surface at $r_{c}$ which means that the slope at $r_{c}$ should be rather small \cite{kipthorne}, i.e.,
\be 
\frac{dz}{dr}_{r=r{c}}=\frac{1}{10},
\ee
which implies,
\be 
r_{c}=10^{4}b_{0}.
\ee 
At $r_{c}\leq r \leq R_{s}$, one chooses, $b(r)=\frac{r}{100}$ and $\Phi(r)=\Phi_{0}$ with $\tau=\rho c^2$ and $p=0$.

For the region $R_{s}\leq r \leq R_{s}+\Delta R$, we have 
\be 
b(r)=\frac{R_{s}}{100}+e^{-2ap^{0}}\frac{(r^{3}-R_{s}^{3})}{3R_{s}^{2}}.
\ee
where we have used $\Delta R=b(R_{s})$ and $\Delta R=\frac{R_{s}}{100}$ in eq.(\ref{transitionlayerb-2}). And for the region $r \geq R_{s}+\Delta R$, we have
\be 
B=b(R_{s}+\Delta R)=\frac{R_{s}}{50}(1-ap^{0}).
\ee
We see that in the transition layer $b(r)$ has NC corrections which was absent in all the analysis of the shape function above.

The redshift function (see Appendix-C for the solution at $r\geq R_{s}+\Delta R$) is given by,
\bea 
\Phi(r)&=&\Phi_{0}~~~~~~~~~~~~~~~~~~~~at~b_{0}\leq r \leq R_{s}+\Delta R\\ \nonumber
\Phi(r)&=&\frac{1}{2}\ln{\left(1-\frac{B}{R}\right)}~~~~~at~r\geq R_{s}+\Delta R.
\eea
Thus we have obtained the complete wormhole solution when the exotic matter is confined to a small radius $r_{c}$ in $\kappa$-deformed space-time.

\end{document}